\documentclass[12pt]{article}
\usepackage{amsmath, amssymb, amsfonts,euscript,mathrsfs}
\usepackage{bm,bbm,latexsym,amsthm}
\usepackage{psfig, epsfig, verbatim}
\usepackage{natbib}
\usepackage{times}
\usepackage[table]{xcolor}
\usepackage{multirow}
\usepackage{xfrac}
\usepackage{color}
\usepackage{colortbl,array} 
\usepackage{multirow,bigdelim}
\usepackage{booktabs}
\usepackage[margin=1in]{geometry}
\usepackage[toc,page]{appendix}
\usepackage{pbox}
\usepackage{subcaption}
\usepackage{ctable}
\usepackage{hyperref}

\newcommand{\PMTwo}{PM$_{2.5}$}
\newcommand{\SOTwo}{SO$_2$}
\newcommand{\NOx}{NO$_x$}
\newcommand{\COTwo}{CO$_2$}

\usepackage{algorithm}
\usepackage{algpseudocode}

\usepackage{setspace}
\newtheorem{assumption}{Assumption}

\title{\Huge\bf Bayesian Nonparametric Trees for Principal Causal Effects}

\begin{document}

\makeatletter 
\newcommand*{\rom}[1]{\expandafter\@slowromancap\romannumeral #1@} 
\makeatother 

\maketitle

\author{\centering {Chanmin Kim$^{1,*}$ Corwin Zigler$^{2}$} \\
$^{1}$Department of Statistics, Sungkyunkwan University, Seoul, Korea\\
$^{2}$Department of Statistics and Data Science, The University of Texas, Austin, TX 78712, USA

{\it *email}: chanmin.kim@skku.edu\\ }
\date{}

\begin{abstract}
Principal stratification analysis evaluates how causal effects of a treatment on a primary outcome vary across strata of units defined by their treatment effect on some intermediate quantity.  This endeavor is substantially challenged when the intermediate variable is continuously scaled and there are infinitely many basic principal strata. We employ a Bayesian nonparametric approach to flexibly evaluate treatment effects across flexibly-modeled principal strata.  The approach uses Bayesian Causal Forests (BCF) to simultaneously specify two Bayesian Additive Regression Tree models; one for the principal stratum membership and one for the outcome, conditional on principal strata.  We show how the capability of BCF for capturing treatment effect heterogeneity is particularly relevant for assessing how treatment effects vary across the surface defined by continuously-scaled principal strata, in addition to other benefits relating to targeted selection and regularization-induced confounding.  The capabilities of the proposed approach are illustrated with a simulation study, and the methodology is deployed to investigate how causal effects of power plant emissions control technologies on ambient particulate pollution vary as a function of the technologies' impact on sulfur dioxide emissions.  
\end{abstract}

\section{Introduction}
Many scientific questions relate to how causal effects of a treatment on a primary outcome relate to the treatment's effect on some intermediate quantity. Complications to causal inference with such intermediate outcomes are well recognized, and have motivated a variety of analysis approaches. In particular, principal stratification \citep{frangakis2002principal} formulates how causal effects of a treatment on an outcome may vary as a function of the treatment's effect on the intermediate quantity. This has proven useful in evaluating causal treatment effects when the intermediate is a surrogate for the primary outcome \citep{gilbert2008evaluating}, a measure of treatment compliance \citep{frumento2012evaluating}, a measure of local response to a national regulation \citep{zigler2018impact}, or time to an event in a study of semicompeting risks \citep{comment2019survivor,nevo2022causal,lyu2023bayesian}. 

The objective of principal stratification is to estimate the average causal effect of a treatment, $A$, on an outcome, $Y$, across ``principal strata'' of the population, where these strata are defined based on the extent to which $A$ causally affects some intermediate variable, $M$ \citep{frangakis2002principal}. Thus, there is a sense in which the presence of treatment effect heterogeneity with respect to the principal strata is a key feature of principal strataification analysis.  Owing to the structure of potential outcomes, which are only observed for each unit under that unit's observed treatment status, units' membership in principal strata are unknown, with allocation of units to these strata a central analysis challenge. When the intermediate $M$ is binary, this amounts to allocating units to one of only four principal strata, and there are numerous available methods \citep{page2012understanding,frumento2012evaluating,miratrix2018bounding}. When $M$ takes on continuous values, there are infinitely many principal strata and the problem is substantially more complex.  Early work in \cite{jin2008principal} offered a Bayesian parametric approach for continuous intermediate variables.  This perspective was extended in \cite{schwartz2011bayesian}, who offered a Bayesian nonparametric Drichlet process mixture (DPM) model for the intermediate outcomes, the primary virtue of which was to estimate a complex joint distribution of potential continuous intermediates that corresponded to very flexibly modeled principal strata. Causal effects conditional on these strata were modeled with a parametric outcome model.  \cite{kim2019bayesian} further extended the flexibility of a DPM to model to both the distributions of potential intermediate and outcome variables, using a Gaussian copula model to combine all marginal distributions into a coherent joint distribution. This approach facilitated estimation of the potentially complex joint distribution of intermediate and outcome variables but, owing to the assumption of normality required in using the Gaussian copula model, the marginal distributions of potential intermediate and outcome variables remained linearly connected, imposing some parametric structure for characterizing principal causal effects. Note that the related framework of causal mediation analysis \citep{pearl2001direct,robins1992identifiability}, with its own Bayesian nonparametric extensions \citep{kim2019bayesian}, is designed to more specifically establish causal mechanism of how a treatment effect is mediated through the intermediate outcome, requiring different types of identifying  assumptions \citep{imai2010identification}.  Relationships between principal stratification and causal mediation analysis are thoroughly explored in \citep{vanderweele2008simple,mealli2012refreshing,kim2019bayesian}, but are not the focus here.   

In this paper, we continue development of flexible Bayesian nonparametric principal stratification analysis with continuous intermediate variables and continuous outcomes. Specifically, we offer an approach that utilizes Bayesian Additive Regression Trees (BART, \cite{chipman2010bart}) to flexibly  model both intermediate and outcome variables. Methods based on BART have been extensively studied in the causal inference community since \cite{hill2011bayesian},  owing to their promising predictive performance. \cite{green2012modeling} used BART in survey experiments to estimate conditional treatment effects. To evaluate heterogeneous treatment effects, \cite{zeldow2019semiparametric} proposed a Bayesian structural mean model (SMM) based on two BART priors. See \cite{tan2019bayesian} and \cite{hill2020bayesian} for a detailed discussion of the BART method and its application in causal inference. Beyond the general goal of incorporating model flexibility and providing a unified estimating procedure for continuous intermediate and outcome variables, we adopt an approach based on BART for reasons specific to the goal of principal stratification.  Since the goal of principal stratification is to evaluate how causal effects of $A$ on $Y$ vary across principal strata, we utilize the BART-derived approach of Bayesian Causal Forests \citep{hahn2020bayesian} for its targeted ability to capture flexible forms of treatment effect heterogeneity, in addition to its ability to adjust for complex confounding structures.  We propose joint model that specifies separate Bayesian causal forest models for continuous intermediate values and primary outcomes, designed specifically to (1) characterize complex principal strata structures and (2) use the capacity of BCF to capture how treatment effects vary across these strata.  A Markov chain Monte Carlo procedure is proposed to sample from the posterior predictive distribution of unobserved potential intermediate variables and estimate principal causal effects.  

The proposed methodology is motivated by and illustrated with an investigation of emission-reduction strategies installed at coal-fired power plants in the United States. Such strategies have played a significant role in the marked reduction of air pollution and corresponding health burden associated with coal power generation in the U.S. \citep{henneman2023mortality}.  The specific treatment of interest is whether a coal-fired electricity generating unit (EGU) has a flue-gas desulfurization scrubber installed to reduce emissions of sulfur dioxide (\SOTwo ) a major byproduct of coal combustion.  Interest lies in the extent to which the causal effect of such scrubbers on ambient fine particulate pollution (\PMTwo, $Y$) vary as a function of the scrubber's impact on \SOTwo\, emissions ($M$).  In particular, the effectiveness of scrubbers in reducing ambient \PMTwo\, may vary with respect to myriad power plant features and circumstances that dictate scrubber impacts on \SOTwo\, such as other changes in operation.  A more complete understanding of the effectiveness of established control strategies, particularly as overall levels of this type of pollution have drastically reduced in the U.S. \citep{henneman2023mortality} is essential and represents a needed advance over similar investigations conducted during earlier time periods \citep{kim2019bayesian, kim2020health}.

\vspace{-0.8cm}
\section{Model and Methods}
\subsection{Notation and Causal Estimands}
The approach is formalized within a potential outcomes framework \citep{rubin1974estimating}.
Let $(A_i, M_i, Y_i)$ denote the exposure, intermediate, and outcome variables of interest, and $\boldsymbol{X}_i$ be a set of confounders for observation $i$, all conceived as fixed pre-treatment features. Throughout the paper, we assume the binary exposure. Let $M_i(\boldsymbol{A})$ denote the potential value of the intermediate variable that would be observed under the vector of exposures $\boldsymbol{A}=(A_1, \ldots, A_n)$ for subjects $i=1, \ldots, n$. Similarly, $Y_i(\boldsymbol{A})$ denotes the potential value of the outcome that would be observed under the vector of exposures $\boldsymbol{A}=(A_1, \ldots, A_n)$. Under the stable unit treatment value assumption (SUTVA, \cite{rubin1980randomization}), potential intermediate and outcome values for subject $i$ do not depend on exposures assigned to other subjects (i.e., $M_i(\boldsymbol{A}) = M_i(A_i)$ and $Y_i(\boldsymbol{A}) = Y_i(A_i)$) and are essentially equal to the observed intermediate and outcome values (i.e., $M_i=M_i(A_i)$ and $Y_i=Y_i(A_i)$). The exposure ($A$) in the motivating air quality study, is the presence of a flue-gas desulfurization scrubber (henceforth, ``scrubber''), measured at each coal-fired power plant in 2014; the intermediate variable ($M$) is the amount of \SOTwo\, emissions from the power plant during 2014; and the outcome of interest is the annual ambient average \PMTwo\, concentration within a 50km radius of the power plant ($Y$). $M_i(A_i)$ and $Y_i(A_i)$ are the potential \SOTwo\, emissions and surrounding \PMTwo\, values based on the state of scrubber installation ($A_i$) at the $i$-th power plant.

Principal stratification \citep{frangakis2002principal} partitions the population based on how $A_i$ affects the intermediate variable, encoded as membership in a principal stratum defined as $S_i = (M_i(0), M_i(1))$. This principal stratum membership $S_i$ is a latent variable that is unaffected by exposure and can be considered the baseline characteristic of the $i$-th subject, encoding features that dictate how the primary treatment impacts the intermediate variable.  In the power plant case, $S_i$ denotes a combination of the effectiveness of scrubbers for reducing \SOTwo\, as well as other plant characteristics dictating variability in this effect that may also relate to the formation of ambient \PMTwo\, such as patterns of operation, responses to regional and seasonal regulatory programs, or atmospheric and weather conditions surrounding the plant.  Our goal is to examine causal effects of $A_i$ on $Y_i$ across principal strata defined by different values of $(M_i(0), M_i(1))$, known as principal causal effects (PCEs, \cite{frangakis2002principal}). 

PCEs can be described as a surface of $\mathbb{E}[Y(1) - Y(0) | M(0), M(1)]$ across values of $(M(0), M(1))$, which has been termed the ``causal effect predictiveness'' surface \citep{gilbert2008evaluating}. Summarizing this surface can be challenging with continuously-scaled $M$, and deriving summary estimands will depend on context.  Expected dissociative effects have been proposed and used to summarize average causal effects of $A$ on $Y$ among principal strata for which $A$ has little or no impact on $M$, representing something similar to a ``direct effect'' of $A$ on $Y$ when $M$ remains unaffected.  Expected associative effects can summarize average effects of $A$ on $Y$ across strata where $M(0) \ne M(1)$.  Here, we extend previous work in \cite{kim2020health} to define multiple average principal effects based on a pre-determined threshold vectors, $\boldsymbol{\epsilon}=(\epsilon_\text{lower},  \epsilon_\text{upper})$  and define expected PCEs as:
\begin{eqnarray*}
\Delta_\text{PCE}(\boldsymbol{\epsilon}) &=& \mathbb{E}[Y_i(1) -Y_i(0) \,|\,  \epsilon_\text{lower} < M_i(1) - M_i(0) < \epsilon_\text{upper}],
\end{eqnarray*}
where various values of $\boldsymbol{\epsilon}$ will define PCEs among strata with different impacts of $A$ on $M$.  For example, specifying $|\epsilon_\text{lower}| = |\epsilon_\text{upper}| \approx 0$ would correspond to an expected dissociative effect among strata of power plants for which scrubbers have little to no effect on \SOTwo\, emissions. If we consider the sample average estimand as the target estimand, the expression for $\Delta_\text{PCE}(\boldsymbol{\epsilon})$ is as follows:
\begin{eqnarray*}
\Delta_\text{PCE}(\boldsymbol{\epsilon}) &=& \frac{1}{n_\epsilon}\sum_{i=1}^n\mathbb{I}\{\epsilon_\text{lower} < M_i(1) - M_i(0) < \epsilon_\text{upper}\}[Y_i(1) -Y_i(0)],
\end{eqnarray*}
where $n_\epsilon = \mathbb{I}\{\sum_{i=1}^n \epsilon_\text{lower} < M_i(1) - M_i(0) < \epsilon_\text{upper}\}$ and $\mathbb{I}\{\}$ is an indicator function.

Note that the fundamental problem of causal inference applies in this case to both the values of $Y_i(A)$ and $M_i(A)$. Thus, the principal stratum membership $S_i$ is unobserved and quantities such as $\Delta_\text{PCE}(\boldsymbol{\epsilon})$ are not identified based only on observed data.  Unobserved values $M_i(1-A_i)$ will determine the value of $S_i$ and must be estimated along with the unobserved values of $Y_i(1-A_i)$. 

\vspace{-0.4cm}
\subsection{Identifying Assumptions}\label{sec:assumptions}
To estimate the principal causal effects, we first formulate an assignment mechanism, the distribution of  $\boldsymbol{A}$ conditional on the potential (intermediate) outcomes $(\boldsymbol{Y}(0), \boldsymbol{Y}(1), \boldsymbol{M}(0), \boldsymbol{M}(1)$) and observed covariates where $\boldsymbol{Y}(a) = (Y_1(a), \ldots, Y_n(a))^\prime$ and $\boldsymbol{M}(a) = (M_1(a), \ldots, M_n(a))^\prime$ for $a=0,1$. Under the assumption that the exposure assignment for each unit is independent of information from other units, the exposure assignment we consider for unit $i$ has the following form: $\text{Pr}(A_i | Y_i(0), Y_i(1), M_i(0), M_i(1), \boldsymbol{X}_i)$.
The ignorability assumption \citep{rosenbaum1983central} states that with a proper set of covariates $\boldsymbol{X}_i$, the exposure assignment mechanism can be reduced to a simpler form.

\begin{assumption} {\bf Strongly ignorable treatment assignment (unconfoundedness)}\label{as:1}
\[\{Y_i(0), Y_i(1), M_i(0), M_i(1)\} \perp \!\!\! \perp A_i \,|\, \boldsymbol{X_i}=\boldsymbol{x}, \]
for all $\boldsymbol{x} \in \mathcal{X}$. Equivalently, it can be restated as $Pr(A_i|Y_i(1), Y_i(0), M_i(1), M_i(0), \boldsymbol{X}_i) = Pr(A_i|\boldsymbol{X}_i)$ where $0<Pr(A_i|\boldsymbol{X}_i)<1$ for $i=1, \ldots, n$.
\end{assumption}
\noindent Here, the ``proper'' set of covariates implies that covariate vector $\boldsymbol{X}_i$ includes all common causes (confounders) of $A-M$ and $A-Y$ relationships. Ignorability implies that, unlike observed $M_i^\text{obs}$, principal stratum membership $S_i = (M_i(0), M_i(1))$ is not affected by the exposure, and average comparisons between $Y_i(0)$ and $Y_i(1)$ within levels of $S_i$ can be interpreted as (principal) causal effects.

Our target estimands are sample average PCEs for the $n$ sample units. Unlike corresponding population average estimands, the parameters governing associations between $Y_i(0)$ and $Y_i(1)$ and $M_i(0)$ and $M_i(1)$ cannot be ignored \citep{schwartz2011bayesian,ding2018causal, kim2019bayesian, li2023bayesian} and we need to apply the notion of inference used in \cite{jin2008principal}.  With the Bayesian causal forest approach described in the next section, we adopt the following conditional independence assumptions to specify two joint conditional distributions of $(Y_i(1), Y_i(0))$ and $(M_i(1), M_i(0))$: $Y_i(1-A_i) \perp Y_i(A_i) | A_i, M_i(0), M_i(1), \boldsymbol{X}_i$ and $M_i(1-A_i) \perp M_i(A_i) | A_i, \boldsymbol{X}_i$. Then, this assumption also further factorizes the joint conditional distributions of potential (intermediate) outcomes as follows:
$Pr(Y_i(0), Y_i(1) |M_i(0), M_i(1),\boldsymbol{X}_i, \boldsymbol{\phi}) = Pr(Y_i(0) |M_i(0), M_i(1),\boldsymbol{X}_i, \boldsymbol{\phi})Pr(Y_i(1) |M_i(0), M_i(1),\boldsymbol{X}_i, \boldsymbol{\phi})$ and 
$Pr(M_i(0), M_i(1) |\boldsymbol{X}_i, \boldsymbol{\phi}) = Pr(M_i(0) |\boldsymbol{X}_i, \boldsymbol{\phi})Pr(M_i(1) |\boldsymbol{X}_i, \boldsymbol{\phi}),$ 
where $A_i$ is omitted in both sides of the equations under the unconfoundedness assumption. This assumption is conservative in that it is stronger than assuming that $(M_i(0), M_i(1))$, and/or $(Y_i(0), Y_i(1))$ are conditionally associated.

\vspace{-0.9cm}
\subsection{Bayesian Causal Forest Overview}
Among many causal inference methods based on BART, \cite{hahn2020bayesian} proposed the Bayesian causal forest model (BCF) to accurately measure heterogeneous effects with the goal of minimizing confounding. Ignoring for the moment the presence of the intermediate variable, $M$, the BCF method specifies the response surface as
\begin{equation}
\mathbb{E}(Y_i\,|\, \boldsymbol{X}_i=\boldsymbol{x}_i, A_i=a_i) = \mu(\boldsymbol{x}_i, \hat{\pi}(\boldsymbol{x}_i)) + \tau(\boldsymbol{x}_i) a_i,\label{bcf}
\end{equation}
where independent BART priors are specified on the functions $\mu$ and $\tau$. Here, $\hat{\pi}(\boldsymbol{x}_i)$ is an estimated propensity score, which corresponds to the exposure assignment mechanism under Assumption 1. 

A key point of the BCF specification is the role of the $\mu$ function, also known as the \emph{prognostic function}, which is equivalent to the expected outcomes of the unexposed (i.e., $\mu(\boldsymbol{x}) = \mathbb{E}(Y(0) | \boldsymbol{x} )$). If the predicted value of the outcome in the absence of exposure relates to the assignment of exposure, the prognostic function will exhibit some dependence on the propensity score $\hat{\pi}(\boldsymbol{x}_i)$. This phenomenon is referred to as \emph{targeted selection} \citep{hahn2020bayesian}. One issue with targeted selection is that it can lead standard regularization methods - including the regularization implied by BART - to present regularization-induced confounding (RIC), particularly when conditional outcome distributions depend more on covariates in $\boldsymbol{X}$ than they do on the treatment, $A$. The inclusion of the estiamted propensity score within the function $\mu$ is designed specifically to mitigate the threat of RIC, and is a key feature of the BCF model in (\ref{bcf}). In the power plant setting, we expect targeted selection to arise because scrubbers are more likely to be installed in power plants emitting a significant amount of pollutants (\SOTwo ) and in areas in danger of violating ambient air quality standards, that is, in areas with the higher values of \PMTwo\, in the absence of emissions controls. 

The other main component of BCF is its inclusion of the function $\tau(\boldsymbol{x}_i)$, which is the feature of the model specifically targeted to a characterization of treatment effect heterogeneity.  Specifically, a flexibly-modeled $\tau(\boldsymbol{x}_i)$ would represent a complex function of covariates that modify the relationship between $A_i$ and the expected outcome.  Estimation in the BCF approach is achieved by specifying an independent BART prior for the $\mu(\boldsymbol{x}_i)$ and $\tau(\boldsymbol{x}_i)$ functions. We provide further details within the context of the proposed model, which incorporates intermediate variables, in a later section.

\vspace{-1cm}
\subsection{Observed Data Model}\label{sec:model}
The observed data models of $Y$ and $M$ can be expressed using the following form of BCF:
\begin{eqnarray*}
Y_i | \boldsymbol{X}_i=\boldsymbol{x}_i, A_i=a_i, M_i(1)=m_{1i}, M_i(0)=m_{0i} & \sim& \mu_Y(\boldsymbol{x}_i, \hat{\pi}(\boldsymbol{x}_i)) + \tau_Y(\boldsymbol{x}_i,m_{1i}, m_{0i}) a_i + \epsilon_i, \\
M_i | \boldsymbol{X}_i=\boldsymbol{x}_i, A_i=a_i &\sim& \mu_M(\boldsymbol{x}_i, \hat{\pi}(\boldsymbol{x}_i)) + \tau_M(\boldsymbol{x}_i) a_i + \epsilon_i^\prime, 
\end{eqnarray*}
where $\epsilon_i \sim N(0, \sigma_y^2)$ and $\epsilon^\prime_i \sim N(0, \sigma_m^2)$. This essentially corresponds to specifying two BCF models; one for the intermediate, $M$, and another for the outcome, $Y$, conditional on both potential intermediate values. These two BCF models correspond to the two joint distributions described as consequences of the conditional independence assumptions described in Section \ref{sec:assumptions}: $Pr(Y_i(0), Y_i(1)|M_i(0), M_i(1), \boldsymbol{X}_i, \boldsymbol{\phi})$, and $Pr(M_i(0), M_i(1)|\boldsymbol{X}_i, \boldsymbol{\phi})$. Here, independent BART priors are specified for the functions $\mu_M$, $\mu_Y$, $\tau_M$ and $\tau_Y$, and $\hat{\pi}(\boldsymbol{x_i})$ represents the estimated propensity score $\hat{\pi}(\boldsymbol{x}_i, \boldsymbol{\phi}_A) = Pr(A_i=1\,|\, \boldsymbol{x}_i, \boldsymbol{\phi}_A)$ for each $i$-th subject where $\boldsymbol{\phi}_A$ is a subvector of the global parameter $\boldsymbol{\phi}$. Key points to highlight include: (1) the flexibility of the outcome and intermediate models in mitigating the threat of RIC while effectively capturing the causal effects of $A$ on both $Y$ and $M$; (2) the reliance of the outcome model on potential intermediate variables $M_i(1)$ and $M_i(0)$ rather than an observed intermediate variable $M_i^\text{obs}$; and (3) the importance of the function $\tau_Y$ in the outcome model, which allows the treatment effect of $A$ on $Y$ to be modified by a complex function of principal stratum membership $S_i = (M_i(0), M_i(1))$ and covariates. The simulation study compares against an approach that excludes $\boldsymbol{X}_i$ from the specification of $\tau_Y$. In either case, estimated PCEs will marginalize over values of $\boldsymbol{X}_i$ within principal strata defined by $S_i$.

By incorporating BART priors, the entire set of observed data models can be reformulated:
\begin{eqnarray*}
Y_i &=& \sum_{h=1}^{H_{\mu_Y}} \mathcal{F}_{\mu_Y} (\boldsymbol{X}_i, \hat{\pi}(\boldsymbol{X}_i, \boldsymbol{\phi}_A); \mathcal{T}_h^{\mu_Y}, \mathcal{M}_h^{\mu_Y}) +  a_i \sum_{h=1}^{H_{\tau_Y}} \mathcal{F}_{\tau_Y} (\boldsymbol{X}_i, M_{i}(1), M_{i}(0); \mathcal{T}_h^{\tau_Y}, \mathcal{M}_h^{\tau_Y}) + \epsilon_i, \\
M_i &=& \sum_{h=1}^{H_{\mu_M}} \mathcal{F}_{\mu_M} (\boldsymbol{X}_i, \hat{\pi}(\boldsymbol{X}_i, \boldsymbol{\phi}_A); \mathcal{T}_h^{\mu_M}, \mathcal{M}_h^{\mu_M}) + a_i \sum_{h=1}^{H_{\tau_M}} \mathcal{F}_{\tau_M} (\boldsymbol{X}_i; \mathcal{T}_h^{\tau_M}, \mathcal{M}_h^{\tau_M}) + \epsilon_i^{\prime},
\end{eqnarray*}
where, for each $f \in \{\mu_Y, \tau_Y, \mu_M, \tau_M\}$, each of $H_f$ distinct tree structures is denoted by $\mathcal{T}_h^{f} (h=1, \cdots, H_f)$  and the parameters at the terminal nodes of the $h$-th tree are denoted by $\mathcal{M}_h^{f} = \{\mu_{h, 1}^f, \cdots, \mu_{h,n_h}^f\}$ where $n_h$ is the number of terminal nodes of $\mathcal{T}_h^f$. And $\mathcal{F}_f(\boldsymbol{x}, \mathcal{T}_h^f, \mathcal{M}_h^f)$ represents $\mu_{t,\eta}^f \in \mathcal{M}_h^f$ if $\boldsymbol{x}$ is associated wtih the $\eta$-th terminal node in tree $\mathcal{T}_h^f$. 

As suggested in \cite{hahn2020bayesian}, we specify different BART priors on $\mu$ and $\tau$ functions. In general, the default parameter setting in \cite{chipman2010bart} is used: (1) a node at depth $d$ splits with the probability $\alpha(1+d)^{-\beta}$ ($\alpha=0.95$ and $\beta=2$); (2) for $\tau_Y$ in the outcome model, independent priors $\mu_{h,j}^{\tau_Y} \sim N(0, \sigma_0^2)$ are assigned to the leaf parameters where $\sigma_0 = \sigma_y / \sqrt{H_{\tau_Y}}$; and (3) for $\tau_M$ in the intermediate model, independent priors $\mu_{h,j}^{\tau_M} \sim N(0, \sigma_0^2)$ are assigned to the leaf parameters where $\sigma_0 = \sigma_m / \sqrt{H_{\tau_M}}$. For $\mu_Y$ and $\mu_M$, however, we specify a weakly informative half-Cauchy prior on the scale parameter of the leaf parameters. We set the prior third quartile to twice the standard deviation of the observed $Y$ (or $M$).

\vspace{-0.8cm}
\section{Estimation}\label{sec:estimation}
\subsection{Posterior Computation}
The detailed Markov chain Monte Carlo (MCMC) approach for posterior computation process is illustrated in the pseudo-code (Algorithm 1) in the Appendix. As shown in Algorithm 1 in the Appendix, a total of four BART priors (two for $M$ model and two for $Y$ model) were updated using ``Bayesian backfitting" \citep{hastie2000bayesian} as a Metropolis-within-Gibbs sampler in which each tree is fit sequentially through the unexplained responses. 
The number of trees included in each BART prior is set differently, as suggested by \cite{hahn2020bayesian}. The BART priors assigned to the prognostic functions $\mu_Y$ and $\mu_M$ that adjust for confounders consist of a large number of trees (e.g., $H_{\mu_Y}=H_{\mu_M}=200$), whereas the BART priors assigned to $\tau_Y$ and $\tau_M$ representing the heterogeneous effects consist of a small number of trees that regularizes towards the homogeneous effects (e.g., $H_{\tau_y}=H_{\tau_m}=50$). 

To summarize the entire procedure, for the $r$-th MCMC iteration, the BART priors assigned to the $\mu_M$ function (prognostic function) and the $\tau_M$ function in the $M$ model are sequentially updated. Thereafter, for each $i=1,...,n$, $M_i^\text{mis}$ is generated, which is an unobservable $M_i(1-A_i)$ value. Through this, the value of the principal stratum membership of the $r$-th MCMC iteration $S_i^{(r)}=(M_i(1), M_i(0))$ is obtained for each $i$-th observation. Similarly, the BART priors corresponding to the $Y$ model are updated. The two potential intermediates, $M_i(A_i)$ and $M_i(1-A_i)$ (forming the principal stratum membership variable, $S_i^{(r)}$), along with $\boldsymbol{X}_i$, serve as independent variables for splitting the tree during the growth of the $\tau_Y$ function.  After updating all of the BART parameters, the $r$-th iteration is completed by updating the variance parameters $(\sigma_m^2, \sigma_y^2)$ with the Gibbs sampler. It is worth noting that the sampling for potential intermediate variables incorporated during the MCMC takes into account the densities for both the $Y$ and $M$ models. Thus, when sampling $M_i^\text{mis}$, it is done conditionally based on the structure of the $Y$ model, which includes the pair of potential intermediate variables. The R code for posterior computation is available at https://github.com/lit777/BPCF.

\vspace{-0.7cm}
\subsection{Causal Effect Estimation}
With $R$ of posterior samples for each parameter and unobserved potential outcome, the posterior means of the principal causal effects for a specific principal stratum $\{S_i: \epsilon_\text{lower} < M_i(1) - M_i(0) < \epsilon_\text{upper}\}$ can be estimated as follows:
\begin{eqnarray*}
\hat{\mathbb{E}}(\Delta_\text{PCE}(\boldsymbol{\epsilon})|\text{Data}) & = & \frac{\sum_{r=1}^R\sum_{i=1}^n \left[Y_i^{(r)}(1) - Y_i^{(r)}(0)\right]  \mathbb{I}\left( \epsilon_\text{lower} <  M_i^{(r)}(1) - M_i^{(r)}(0) < \epsilon_\text{upper}\right)}{\sum_{r=1}^R \sum_{i=1}^n \mathbb{I}\left( \epsilon_\text{lower} <  M_i^{(r)}(1) - M_i^{(r)}(0) < \epsilon_\text{upper}\right)},
\end{eqnarray*}
where $(M_i^{(r)}(0), M_i^{(r)}(1))$ and $(Y_i^{(r)}(0), Y_i^{(r)}(1))$ are the $r$-th MCMC samples of potential intermediate and outcome variables, respectively. 

Multiple values of $\boldsymbol{\epsilon}$ can be specified to investigate how the effect of $A$ on $Y$ varies as a function of $A$ on $M$ by analyzing the outcome effects within groups categorized based on the strength of the exposure-intermediate relationship. For example, a sequence of $\Delta_\text{PCE}(\boldsymbol{\epsilon})$ estimate to increase in magnitude as $|\epsilon_\text{upper} - \epsilon_\text{lower}|$ grows would indicate that larger effects on the intermediate are associated with larger impacts on the outcome.  We will motivate and explore a series of $\Delta_\text{PCE}(\boldsymbol{\epsilon})$ in the context of the analysis of power plant data.

\section{Simulation}
We examine the model performance based on simulated datasets with $N=300$ observations. We consider two different scenarios: (Scenario I) seven confounders ($X_1-X_7$) are independently generated from $N(0,1)$. $Y$, $M$, and $A$ are confounded through $X_1- X_7$; (Scenario II) two confounders ($X_1-X_2$) are independently generated from $Unif(0, 1)$ and three confounders are independently generated from normal distributions. $Y$, $M$ and $A$ are confounded through $X_1- X_5$. The second scenario represents the targeted selection situation where individuals select into exposure based on a prediction of the unexposed outcome \citep{hahn2020bayesian}. The data generating process and results for Scenario II are in the Appendix. In both scenarios, confounding is strong and effect heterogeneity exists through interactions between $A-X$ in the $M$ model and between $A-S$ in the $Y$ model. We generate $m=200$ simulated data under each scenario. The MCMC chain runs for 10,000 iterations, and the first 5,000 iterations are discarded as burn-in. 

The number of trees in our proposed model (denoted by the Bayesian principal causal forest; BPCF) was set to 150 for the prognostic function of the intermediate/outcome models, and 50 for the modifier function in the intermediate/outcome models. The prognostic and modifier functions of the intermediate model contain the same set of covariates, whereas the outcome model's modifier function contains the principal stratum membership $S_i = (M_i(0), M_i(1))$ along with the full set of covariates. A node at depth $d$ splits with the probability $\alpha(1+d)^{-\beta}(\alpha = 0.95\text{ and } \beta= 2)$ for the prognostic function but with the probability $\alpha(1 + d)^{-\beta}( \alpha = 0.25\text{ and }\beta= 3)$ for the modifier function to shrink more towards homogeneous effects. To compare our model performance to other methods, we consider three competing models: (1) seperate BART models for $\{M(0), M(1),Y(0),Y(1)\}$, which is denoted by `BART$_\text{pce}$'; (2) Dirichlet process mixture (DPM) model from \cite{schwartz2011bayesian}, which is denoted by `DP$_\text{pce}$'; and (3) the proposed BPCF without covariates $\boldsymbol{X}$ in the modifier function of the outcome model (i.e., the modifier function $\tau_Y$ only contains $S_i$), which is denoted by `BPCF$_\text{m only}$'. The specific forms of the BART$_\text{pce}$ and DP$_\text{pce}$ models are included in the Appendix.

\vspace{-0.8cm}
\subsection{Simulation I}

For the first simulation scenario, we consider the following data generating process:
\begin{eqnarray*}
Y_i &=&  h_1(x_{i1}) + 1.5h_2(x_{i2})- (M_i(1) - M_i(0))^2 \times A_i +2 |x_{i3}+1| + 2x_{i4} \\
& & + \exp(0.5 x_{i5}) -0.5  |x_{i6}| -  |x_{i7}+1|  + \epsilon_i,\\
M_i & = & 0.5 h_1(x_{i1})+0.5 h_2(x_{i2})+2 A_i + |x_{i3}+1| + 1.5 x_{i4}\\
& &  - \exp(0.3 x_{i5}) +  A_i\times |x_{i5}| + \psi_i,\\
\mathbb{E}(A_i | \boldsymbol{x}_i) & = &  \Phi\left(0.5 + h_1(x_{i1}) + h_2(x_{i2})-0.5 |x_{i3}-1|+1.5 x_{i4} \times x_{i5}  \right),\\
\epsilon_i &\sim & N(0,0.3^2),  \,\,\, \psi_i \sim N(0, 0.1^2), \,\,\, x_{ij} \sim N(0,1) \text{  for  } j=1, \ldots, 7,
\end{eqnarray*}
where $h_1(x) = (-1)^{I(x\geq 0)}$ and $h_2(x) = (-1)^{I(x < 0)}$. This simulation scenario is developed to evaluate the accuracy of PCE estimation when the $Y$ and $M$ models are given a complex structure. 

In this simulation study, target PCEs were defined based on five different intervals. These intervals were determined using quintiles of observed intermediate values obtained through the data generating process. Specifically, the first interval was (2.00, 2.26), the second interval was (2.26, 2.53), the third interval was (2.53, 2.84), and the fourth interval was (2.84, 3.29), and the last interval was (3.29, 5.09).
The results are shown in Table \ref{tab:sim1}. Relative bias and MSE were compared, and the proposed BPCF performed relatively better across all PCEs. In particular, it demonstrated very high accuracy in estimating the causal effect for intermediate variables $[M_i(1) - M_i(0)]$, and it can be argued that it showed high performance in estimating PCEs by estimating accurate principal stratum membership based on precise $[M_i(1) - M_i(0)]$ estimates.

\begin{figure}[hp]
\begin{subfigure}{.5\textwidth}
  \centering
    \caption{Truth}
  \includegraphics[width=.95\linewidth]{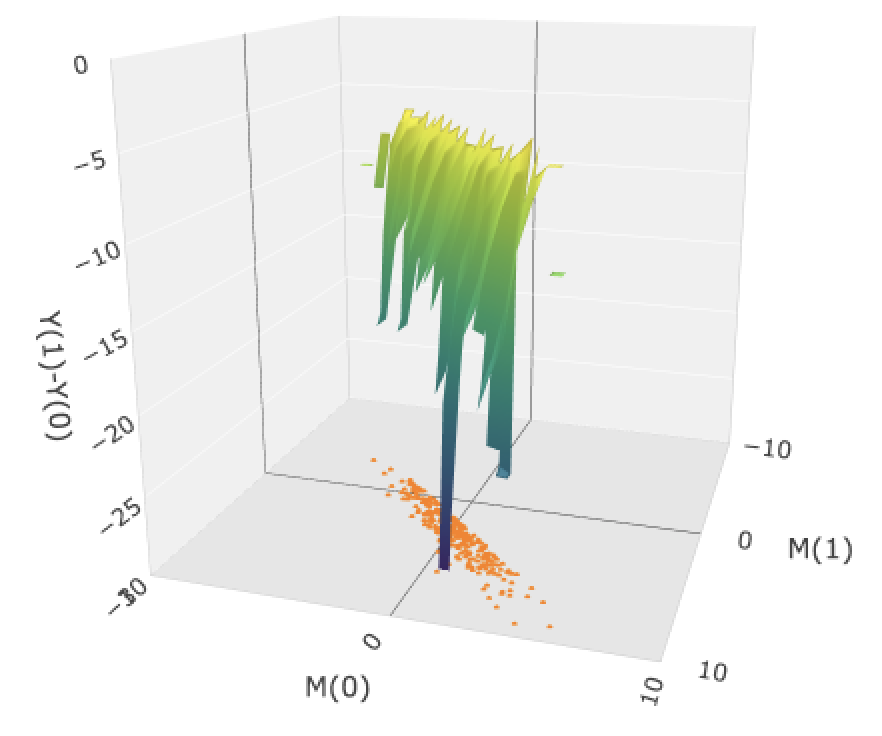}
  \label{fig2:sfig1}
\end{subfigure}%
\begin{subfigure}{.5\textwidth}
  \centering
    \caption{BPCF}
  \includegraphics[width=.95\linewidth]{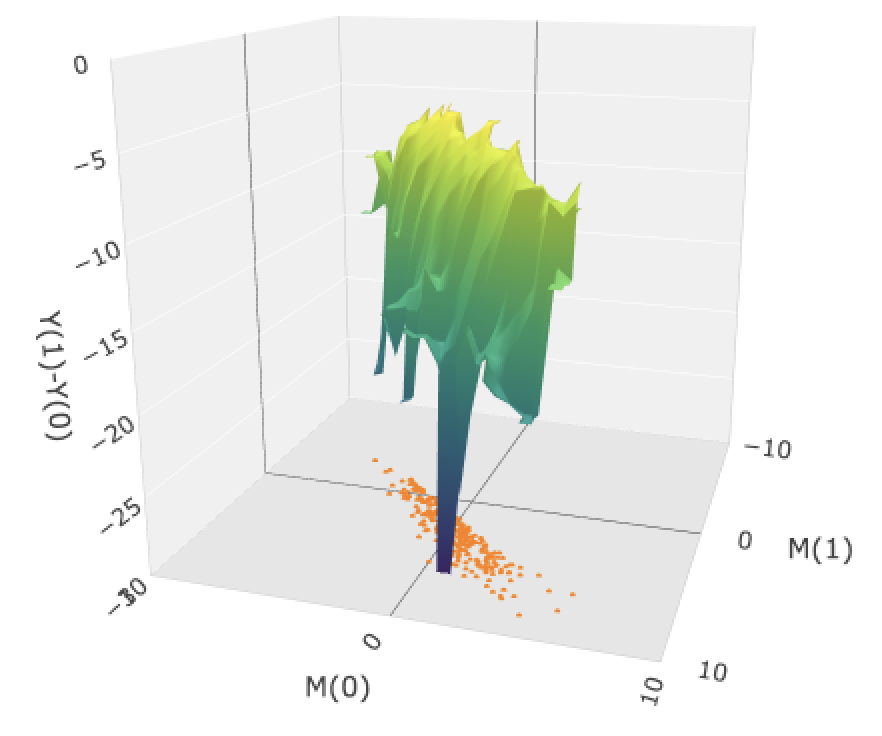}
  \label{fig2:sfig2}
\end{subfigure}
\begin{subfigure}{.5\textwidth}
  \centering
    \caption{BPCF$_\text{m only}$}
  \includegraphics[width=.95\linewidth]{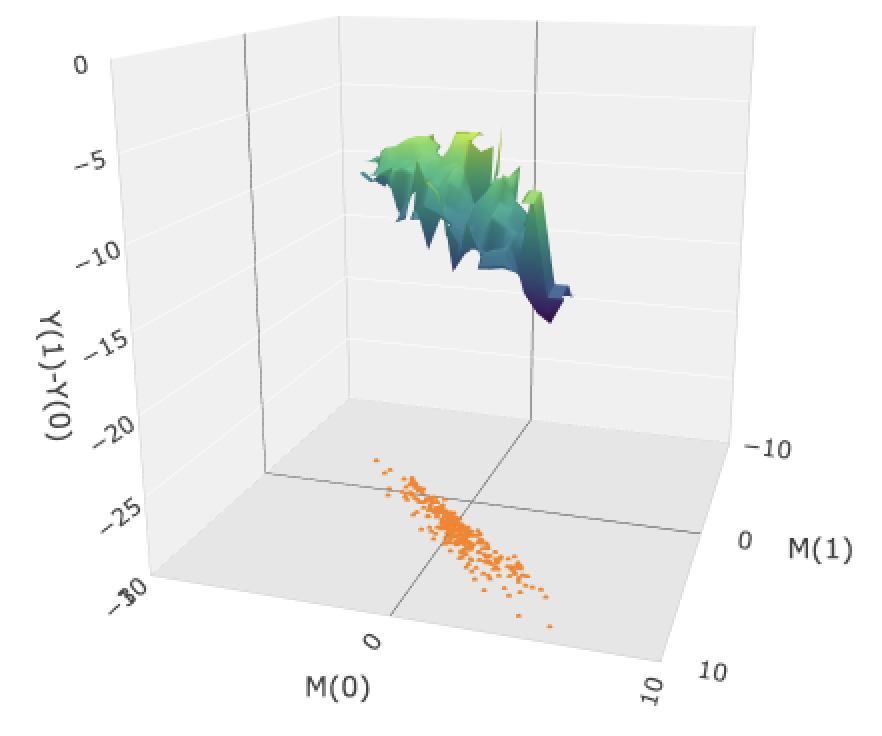}
  \label{fig2:sfig2}
\end{subfigure}
\begin{subfigure}{.5\textwidth}
  \centering
    \caption{BART$_\text{pce}$}
  \includegraphics[width=.95\linewidth]{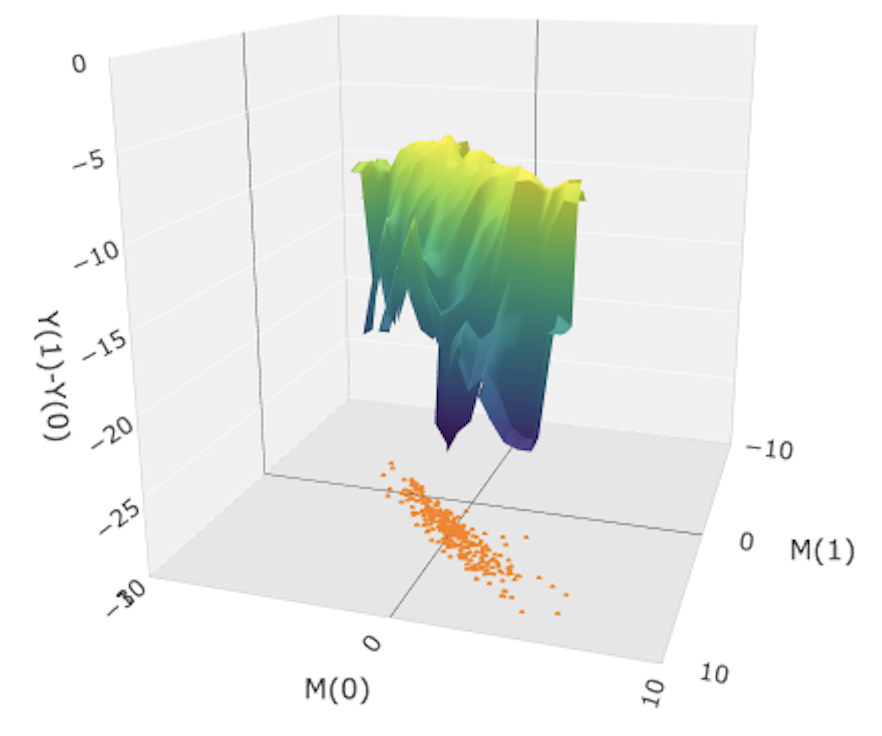}
  \label{fig2:sfig3}
\end{subfigure}
\begin{subfigure}{.5\textwidth}
  \centering
    \caption{DP$_\text{pce}$}
  \includegraphics[width=.95\linewidth]{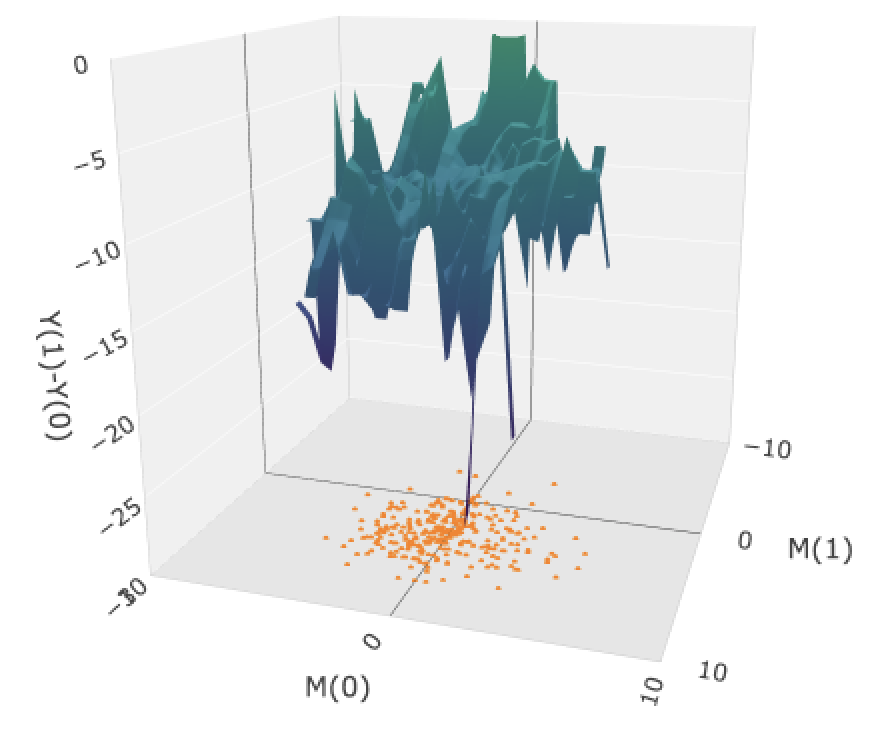}
  \label{fig2:sfig4}
\end{subfigure}
\caption{[{\bf Simulation I}] Average surface plots for all combinations of $(M(0), M(1))$ from four different models: (a) Truth, (b) BPCF, (c) BPCF$_\text{m only}$, (d) BART$_\text{pce}$, and (e) DP$_\text{pce}$. The points in $X-Y$ plane are averages of one set of MCMC samples.  As a response surface, the corresponding value of the effect $A$ on the outcome $E[Y(1)-Y(0)]$ is plotted.}
\label{fig:sim1}
\end{figure}

A causal effect predictiveness surface \citep{gilbert2008evaluating} was created to help interpret the results of each model more precisely. Figure \ref{fig:sim1} depicts the predictiveness surfaces created using four different methods, including true surface plots. The $x-y$ plane has two axes representing $M(1)$ and $M(0)$ variables, and the $z$-axis represents the corresponding $Y(1)-Y(0)$ value. In terms of the overall response slope, the DP method differed significantly from the actual shape. The dots on the $x-y$ plane represent $M(1)$ and $M(0)$ sample values from one MCMC iteration. Based on this, it is reasonable to interpret that the BART and BPCF methods form more accurate principal strata than does DP.
 In particular, it can be observed that the DP method failed to accurately estimate $S_i=(M_i(1), M_i(0))$. The reason for this phenomenon occurring is that the data generating model for $M$ has a complex nonlinear structure that includes the interaction of $A$ and $\boldsymbol{X}$. In contrast, the DP method (based on \cite{schwartz2011bayesian})  is designed to vary only the intercept parameters of $M(1)$ and $M(0)$. Extensions of \cite{schwartz2011bayesian} to extend the DP model to additional model parameters might improve performance in a scenario where interactions between $A$ and $\boldsymbol{X}$ dominate the data generating model for $M$.  On the other hand, the BPCF$_\text{m only}$ method relatively accurately estimated the values of $S_i$, but the response surface showed deviations from the true values. This can be interpreted as the tree structure shrinking towards a more homogeneous effect when the modifier function only includes $S_i$ and not all covariates.

\begin{table}[ht]
\centering
\caption{Simulation results (relative bias (rBias) and mean square error(MSE)) under Scenario I} \label{tab:sim1}
\resizebox{.99\textwidth}{!}{%
\begin{tabular}{|c|c|cc|cc|cc|cc|}
\hline
\multirow{2}{*}{Category} & \multirow{2}{*}{True Effect} & \multicolumn{2}{c|}{BPCF} & \multicolumn{2}{c|}{BPCF$_\text{m only}$} & \multicolumn{2}{c|}{BART$_\text{pce}$} & \multicolumn{2}{c|}{DP$_\text{pce}$}  \\
\cline{3-10}
& & rBias & MSE & rBias & MSE & rBias & MSE & rBias & MSE   \\
\hline
$E[M(1)-M(0)]$ & 2.79 & -0.002 & 0.001 & 0.001 & 0.001 & -0.006 & 0.001 & -0.014 & 0.0014\\
$E[Y(1)-Y(0)| \text{ interval 1}]$ & -4.54 & 0.14 &  0.51 & 0.40 & 3.66 & 0.30 &  1.94& 0.63 & 11.25\\
$E[Y(1)-Y(0)| \text{ interval 2}]$& -5.71 & 0.02 & 0.09 & 0.19 & 1.39 & 0.09 & 0.29& 0.26& 3.84\\
$E[Y(1)-Y(0)| \text{ interval 3}]$ & -7.18 & -0.03 & 0.13 & 0.04 & 0.31 &  -0.05 & 0.19 & -0.02& 0.74\\
$E[Y(1)-Y(0)| \text{ interval 4}]$ & -9.28 & -0.06 & 0.40 & -0.09 & 1.16 &  -0.14 & 1.86 & -0.24& 5.39\\
$E[Y(1)-Y(0)| \text{ interval 5}]$ & -14.11 & -0.10 & 2.42 & -0.27 & 15.36 &  -0.23 & 10.47 & -0.48& 48.35\\
\hline
\end{tabular} %
}
\end{table}

\section{Evaluation of Power Plant Scrubber Impacts on Emissions and Ambient Air Quality}\label{sec:application}
The plan to reduce US \SOTwo\, emissions to 1980 levels through the Acid Rain Program established by Title IV of the Clean Air Act achieved its goal primarily by reducing emissions from fossil fuel power plants, or, more formally, electricity-generating units (EGUs). The total estimated annualized human health benefits from ARP range from \$50 billion to \$100 billion \citep{chestnut2005fresh}, but these estimates are heavily reliant on assumed relationships, particularly those between the program, power plant emissions, and ambient \PMTwo. A recent study \citep{kim2019bayesian} discovered a significant relationship among them using data from 2005. Recent data-driven studies, on the other hand, are scarce.

In \cite{kim2019bayesian}, the impact of emission-control technology on ambient air quality was analyzed using a flexible statistical model that considered multiple pollutant emissions. By estimating the direct and indirect effect of scrubbers on  \PMTwo\, a significant associative effect associated with \SOTwo\, emissions was determined. It is worth noting that these findings were documented in 2005, and no subsequent studies extending to time periods where this type of pollution was comparatively low have been conducted. Additionally, the study had limited consideration for other factors, such as social and environmental factors, that may influence the selection of a scrubber among various methods for \SOTwo\, reduction.

\subsection{Data}
We collected data on 385 coal fuel power plants located in the eastern United States and operating during 2014, incorporating information from the EPA Air Markets Program Data and the Energy Information Administration. The data encompassed plant characteristics, installed emissions control technologies (if applicable), and emissions of various pollutants including \SOTwo. Additionally, predictions of annual \PMTwo\, concentrations at the zip code level were obtained \citep{weidata}. Our analysis specifically centered around the year 2014.
In terms of confounders, weather variables are major factors that can affect the operation of power plants and the concentration of ambient \PMTwo. We characterized meteorological conditions at the zip code level by measuring relative humidity, temperature, and precipitation at the zip code's centroid \citep{tec2022weather2vec}. Non-local meteorological information from surrounding areas is also a crucial factor that affects local air quality. Recent work in \cite{tec2022weather2vec}, proposed the use of self-supervised learned latent representations of non-local meteorological information surrounding each zip code as important confounding variables in studies of air pollution. We use the learned representations of non-local meteorological variables within a 100km radius of each zip code as additional confounders which can be extracted from \href{https://huggingface.co/spaces/mauriciogtec/w2vec-app}{https://huggingface.co/spaces/mauriciogtec/w2vec-app}. Furthermore, the 2010 US Census data provided critical demographic information (i.e., total population) for a potential confounder surrounding each power plant. 

Using the centroids of the zip codes, we established connections between each power plant and all zip code locations within a specified radius. This approach enabled us to link the average ambient \PMTwo\, concentration and the average value of local confounders to each power plant. We used a 50km radius in the main manuscript, with an additional radius (60km) being examined in the Appendix. 
As of 2014, a power plant was categorized as treated if over half of its EGUs had implemented any control techniques for sulfur dioxide (\SOTwo). Additionally, only power plants with a total heat input greater than 0 were considered, and power plants with missing values for important confounding factors (especially power plant characteristics) were excluded. 
Table \ref{Data} shows the summary statistics for each variable in the two treatment groups. Before presenting the analytical results, MCMC convergence was assessed with plots of the MCMC log-marginal likelihood (Figure S2 in the Appendix).

\begin{table}[h]
\centering
\caption{Summary statistics for covariates, intermediate and outcomes available for the analysis. All variables represent annual averages, with Census variables relying on annual data starting from the year 2010. Note that there are also 10 additional learned latent representations of neighboring weather information included in the analysis. }\resizebox{\textwidth}{!}{  
\begin{tabular}{lcccc} \hline \hline
 & \multicolumn{2}{c}{\underline{Have \SOTwo\, control (n=74)}} & \multicolumn{2}{c}{\underline{Have no \SOTwo\, control (n=87)}} \\
  & Mean & SD & Mean & SD \\ \hline
   \underline{Air Quality and Weather Data} \\
Ambient PM$_{2.5}$ ($\mu\text{m}$)  & 9.15 & (1.70) & 8.89 & (1.64)\\
Relative Humidity(\%) & 74.32 & (3.88) & 73.95  & (5.00))\\
Temperature at 2m($^\circ C$) & 11.86 & (3.49) & 11.62 & (4.10)\\
Total Precipitation($kg/m^2$) & 2.73 & (0.52) & 2.70 & (0.49)\\

 \\ \underline{Power Plant Data} \\
Total \SOTwo\, Emission (tons)& 713.95 & (818.80) & 1063.70 & (1353.82)\\
Total \NOx\, Emission (tons; past year) & 542.47 & (400.97) & 365.96 & (365.69)\\ 
Total \COTwo\, Emission (1000 tons; past year) & 704.16 & (424.70) & 334.89 & (381.71)\\ 
Average Heat Input (10000 MMBtu)& 8139.89& (4825.14) & 4446.26 & (3870.44) \\
Phase of Participation in the ARP (Phase I or II) &0.62& (0.48) & 0.75 & (0.44)\\
Total Operating Time (hours $\times$ \# units) &19329.67& (9843.08) & 18464.60& (10605.52) \\
Sulfur Content in Coal (lb/MMBtu) &2.09& (1.23) &0.89& (0.96)\\
Num. of \NOx\, Controls (\# units) & 4.97 & (2.93) & 4.09 & (2.45)\\
Num. of Units & 2.85 & (1.49) & 3.31 & (1.71) \\
Gross Load (1000MWh) & 8455.37 & (5042.04) & 4603.13 & (4111.47) \\
Capacity (\%) & 58.22 & (11.55) & 47.48 & (16.68) \\

 \\ \underline{Census Data} \\
Population 2010 & 8604.69 & (6612.86) & 10077.72 & (7377.57)\\

\end{tabular}}
\label{Data}
\end{table}

\subsection{Preliminary Analysis}
A simple average comparison of the two groups showed that the concentration of ambient \PMTwo\, was approximately 0.25 ($\mu\text{m}/\text{m}^3$) higher within 50 km of the power plants where \SOTwo\, control technologies were installed. The average \SOTwo\, pollution emissions from the power plants with and without any \SOTwo\, control techniques were 713.95 tons and 1063.70 tons, respectively, indicating that the group of power plants without \SOTwo\, control technologies emitted more (roughly 349.76 tons more) \SOTwo\, pollution. Propensity score estimates were obtained using logistic regression with the confounding variables mentioned in the previous section.

Subsequent estimates are reported as posterior means along with 95\% credible intervals. The estimated causal effect of \SOTwo\, control strategies on \SOTwo\, emissions was -480.09 (-702.55, -378.54) tons, indicating a significant reduction in \SOTwo\, emissions. Similarly, the estimated causal effect of control strategies on ambient \PMTwo\, concentrations was -0.60 (-0.83, -0.37), showing a significant reduction in \PMTwo\, concentrations within a 50-kilometer radius.
These results coincide with the expectation that introducing \SOTwo\, reduction technology in power plants would consistently result in reduced \SOTwo\, emissions and subsequently lower ambient \PMTwo\, levels. The principal stratification approach will more thoroughly examine the influence of power plants' varying response in \SOTwo\, emissions on the downstream impacts on ambient \PMTwo\,.

\begin{figure}[h]
\centering
\includegraphics[width=15cm]{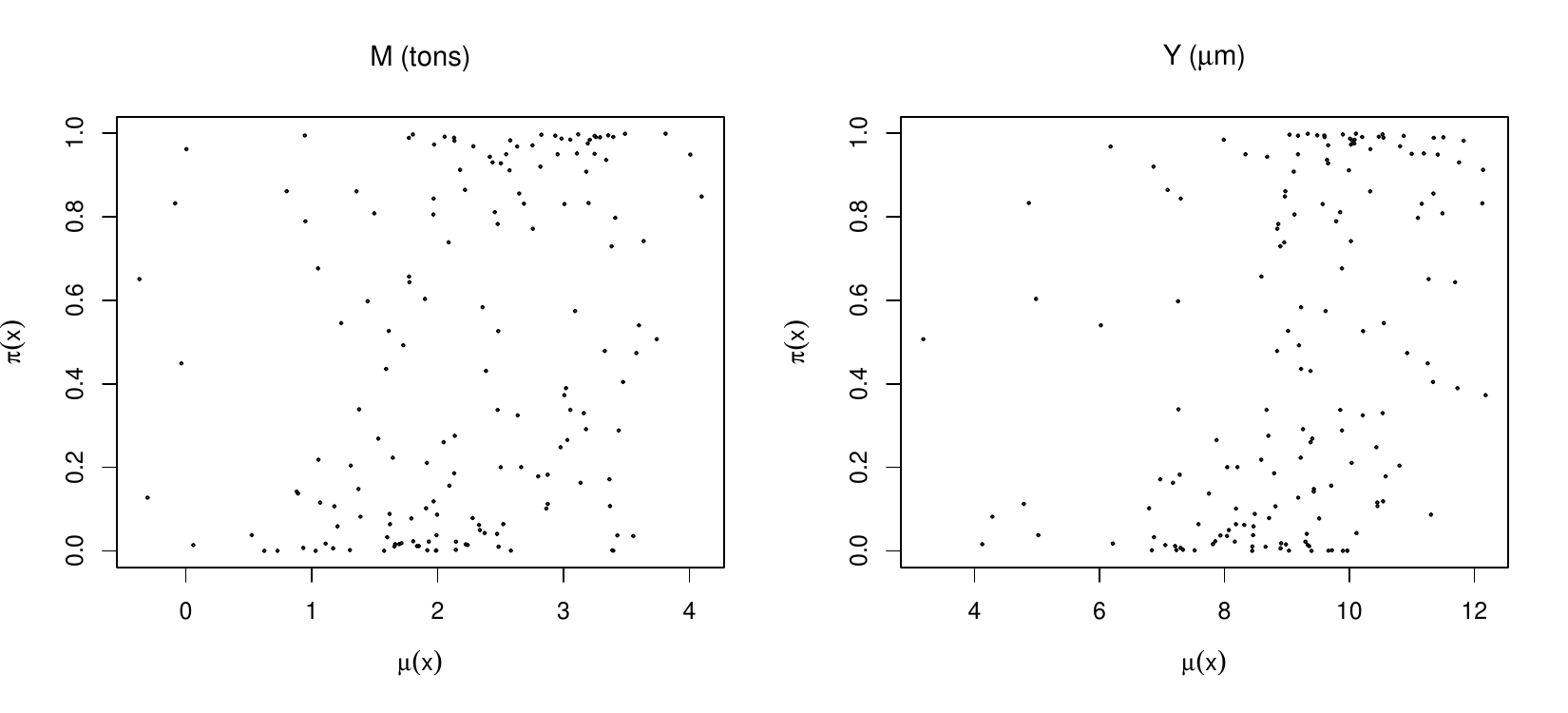}
\caption{For the estimated prognostic values $\mu(\boldsymbol{x}_i)$, the estimated propensity scores $\pi(\boldsymbol{x}_i)$ are plotted. The right plot for \PMTwo\, concentrations ($Y$) demonstrates a moderate monotone relationship between the propensity score and the prognostic function.} \label{fig:RIC}
\end{figure}

Notably, Figure \ref{fig:RIC} indicates that, in both the intermediate and outcome models, there is a positive relationship between propensity score estimates and prognostic function estimates. Particularly for the outcome variable (\PMTwo), an increasing relationship between $\mu$ and $\pi$ indicates that targeted selection bias may exist. This serves as one key motivation for the BCF-based approach.

\begin{figure}[h]
\centering
\includegraphics[width=17cm]{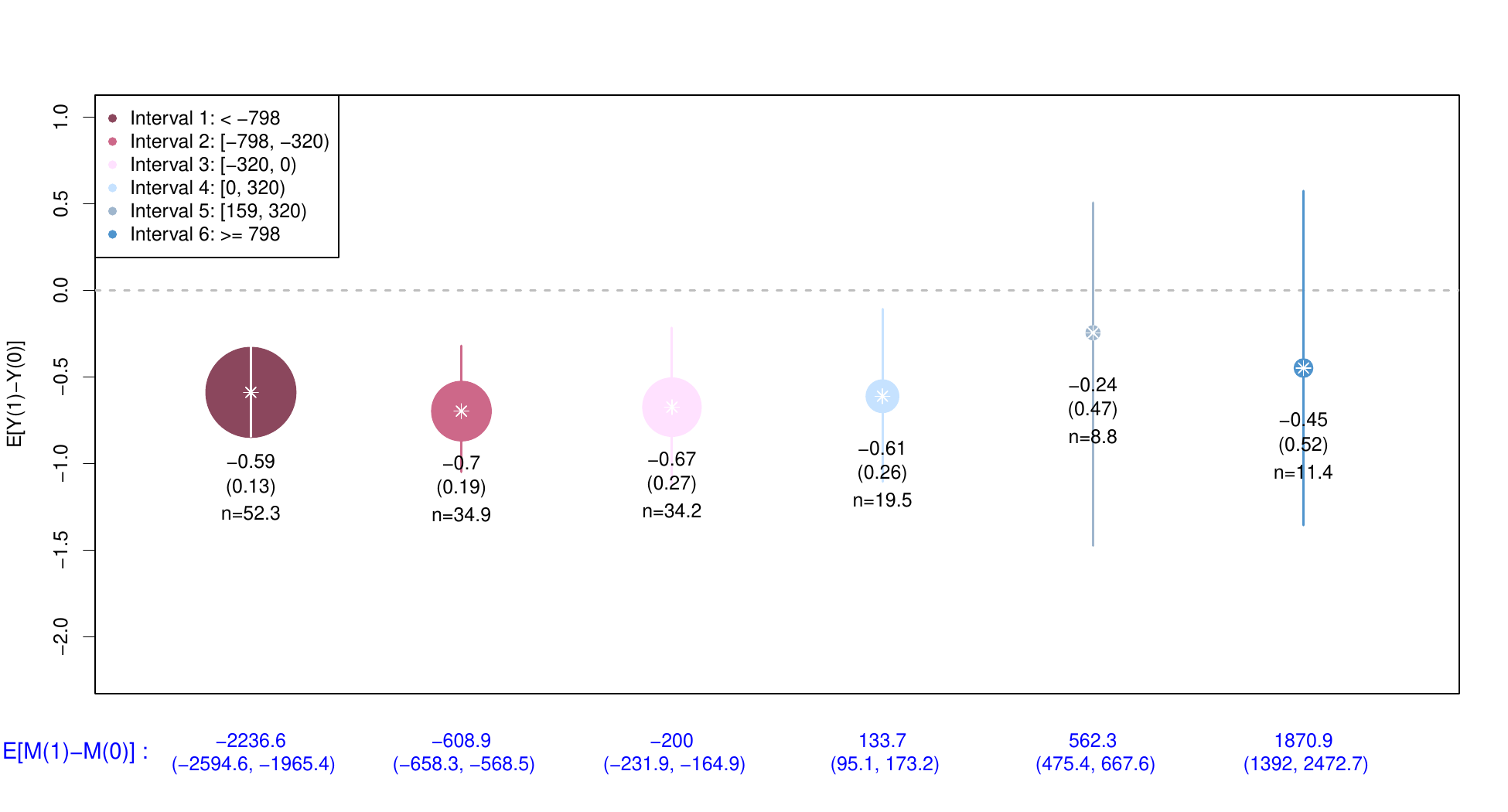}
\caption{Posterior mean estimates of principal causal effects for six different strata. These six strata are defined by intervals 1, 2, 3, 4, 5, and 6. The size of the circle is proportional to the number of plants in that stratum, and numbers listed are posterior mean of the effects, standard deviation and the number of subjects, respectively. } \label{fig:main}
\end{figure}

\subsection{Principal Causal Effects}

Principal strata chosen for summarizing the CEP surface were constructed based on the estimated variability among potential intermediate variables. The posterior predictive mean values for each $i$-th $M_i(1)-M_i(0)$ was estimated, and multiplying the standard deviation of these values by 0.2 and 0.5 resulted in values of 320 and 798, respectively. Using this information, six principal strata were constructed as follows.: interval 1 is ($<$ -798), interval 2 is [-798, -320), interval 3 is [-320, 0), interval 4 is [0, 320), interval 5 is [320, 798), and interval 6 is ($\geq$ 798). The strata defined by the first three intervals demonstrate how the \PMTwo\, concentrations vary within units where \SOTwo\, emissions decrease due to \SOTwo\, control installation. The last three intervals explain the opposite impact of \SOTwo\, control techniques on \PMTwo\, concentrations in situations where \SOTwo\, emissions actually increase.

In the case of the principal causal effect for interval 1 ($\Delta_{\text{PCE} 1}$), 52.3 power plants were assigned to the subgroup on average, and the estimated value of the causal effect was -0.59 (95\% CI:  -0.84, -0.31). In the case of the principal causal effect for interval 2 ($\Delta_{\text{PCE} 2}$), 34.9 power plants were assigned to the subgroup on average, and the estimated value of the causal effect was -0.69 (95\% CI: -1.05, -0.32). For interval 3($\Delta_{\text{PCE} 3}$), 34.2 power plants were assigned to the subgroup on average, and the estimated value of the causal effect was -0.67 (95\% CI: -1.13, -0.22). On the other hand, in the case of the principal causal effects for intervals 4-6 ($\Delta_{\text{PCE} 4}$,$\Delta_{\text{PCE} 5}$,$\Delta_{\text{PCE} 6}$), the groups with increased \SOTwo\, emissions had averages of 19.5, 8.8, 11.4 power plants and estimated effects of -0.61 (95\% CI: -1.10, -0.11), -0.24 (95\% CI: -1.47, 0.51), -0.45 (95\% CI: -1.36, 0.57), respectively. The first three PCEs displayed significant effect estimates, with more power plants assigned to those strata. This suggests that the implementation of \SOTwo\, control techniques led to a substantial reduction in \SOTwo\, emissions and consequently lowered \PMTwo\, concentrations. Specifically, based on the fact that the highest number of power plants was assigned to the first principal stratum, it can be inferred that a significant number of power plants achieved a reduction in \SOTwo\, emissions of 798 tons or more through the installation of \SOTwo\, control techniques. In contrast, the last three principal strata, indicating a causal increase in \PMTwo\, concentration due to the installation of \SOTwo\, control, did not show significant PCEs except for interval 4. Moreover, the number of power plants within these strata was observed to be relatively small. This suggests that the installation of \SOTwo\, controls tends to not increase \SOTwo\, emissions and, furthermore, exhibits little evidence of an impact on \PMTwo\, concentrations.

\begin{figure}[h]
\centering
\includegraphics[width=11cm]{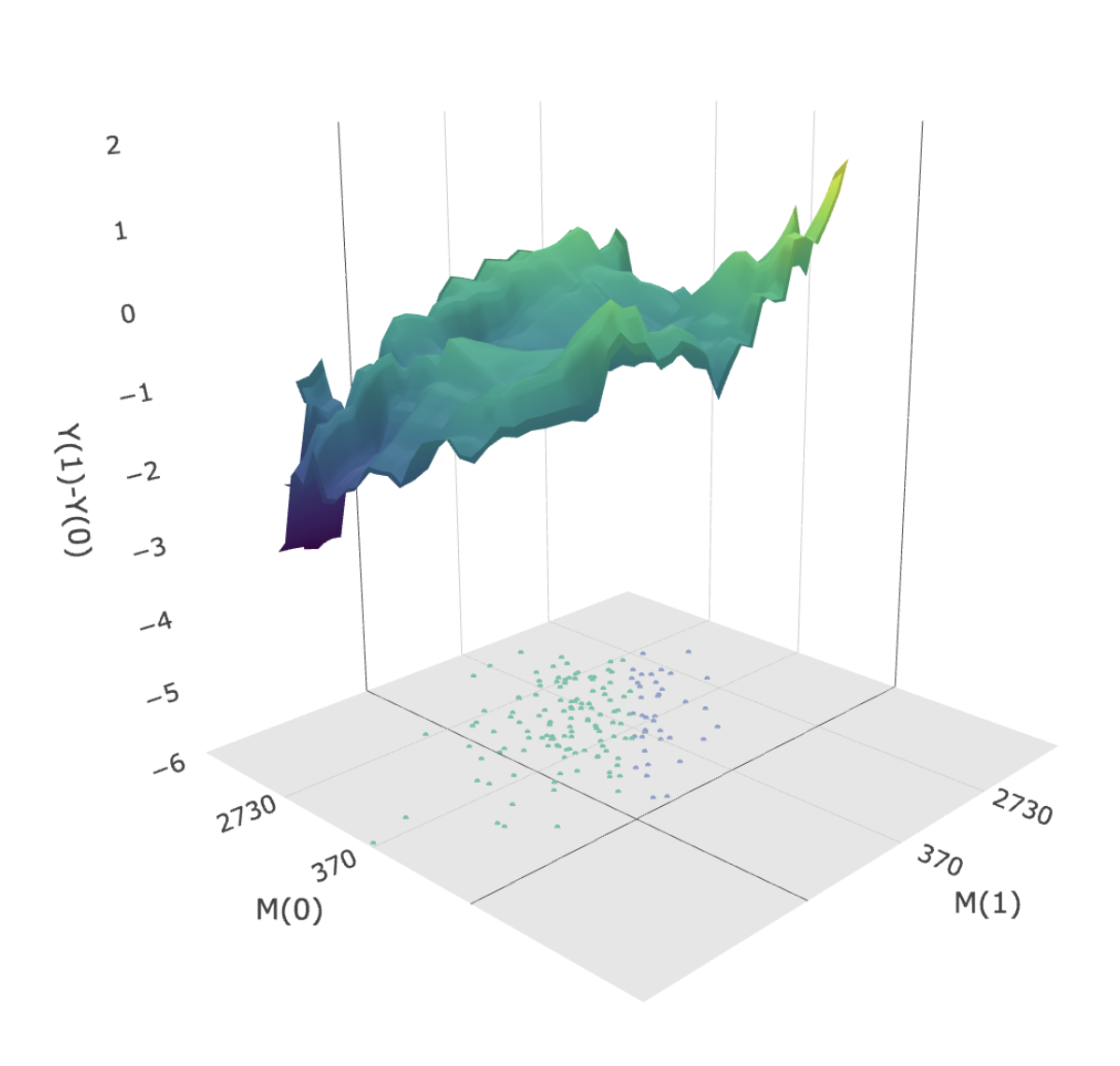}
\caption{Average surface plot of the causal effect on \PMTwo\, for different values of $(M(0), M(1))$ by tons. Posterior predictive values of $(M(0), M(1))$ are plotted on the $x-y$ plane. The corresponding value of the causal effect of \SOTwo\, control techniques installation on \PMTwo, $E[Y(1)-Y(0)]$, is plotted on the $z-$axis.}\label{fig:response}
\end{figure}

These results are well illustrated in Figure \ref{fig:main} and Figure \ref{fig:response}. Figure \ref{fig:main} displays the principal causal effects corresponding to intervals 1-6 from left to right. The center point of each circle represents the posterior mean of the principal causal effect, and the vertical line represents the 95\% credible interval. The size of each circle represents the average value of subjects within the respective stratum, and the numbers below indicate the posterior mean, posterior standard deviation (SD), and average number of subjects in sequential order. The $x$-axis represents the causal effect on \SOTwo\, emissions within each interval.

Figure \ref{fig:response} presents the average response surface. The points on the $x-y$ axis represent the posterior predictive mean values for $S_i=(M_i(1), M_i(0))$. It can be observed that the slope of the response surface generally decreases in regions where $M(1)$ values are lower than $M(0)$ values. This indicates that the reduction in \SOTwo\, emissions corresponds to a decrease in \PMTwo\, concentrations, with steeper impacts in areas of the largest \SOTwo\, reductions.

\subsubsection{Additional Analysis}
In the main analysis, each power plant utilized the average \PMTwo\, concentrations of zip codes within a 50km radius of the zipcode centroid as the outcome variable. However, considering the atmospheric movement, the area directly influenced by each power plant's air quality may be different. Therefore, an additional analysis was conducted by obtaining the average \PMTwo\, concentration within a 60km  radius and utilizing them as outcome variables. This range is set to prevent excessive overlap between each power plant's sphere of influence. The appendix provides detailed information on the the result. When comparing these results to those within a 50km radius, it was confirmed that there is not a significant difference between the two sets of results.

\section{Discussion}
This paper presented a novel principal causal effect estimation method that can investigate the role of intermediate variables in estimating causal effects. To estimate the principal causal effect, which varies with changes in principal stratum membership, intermediate, and outcome variable models were created using the Bayesian causal forest model, which is known to accurately estimate the heterogeneous effect. This resulted in a more detailed characterization of principal strata, which were plugged into the modifier function in the outcome model as an effect modifier. The proposed method provides protection against the evident threat of targeted selection and regularization-induced confounding. The air quality study in Section 5 demonstrated the phenomenon of targeted selection in relation to the effect of \SOTwo\, emission reduction technology on \PMTwo\, concentrations, proving the efficacy of the proposed method.

In this study, a priori estimated propensity scores were utilized as fixed covariates, which means that the uncertainty surrounding the propensity score estimation was not considered in the final analysis. As claimed in \cite{hahn2020bayesian}, incorporating propensity score estimates is merely a means to control regularization by transforming covariates. Therefore, a priori estimation is not a major concern. Jointly modeling the propensity score model with the intermediate and outcome models is not a straightforward task, primarily due to the feedback between these models. To address this limitation and avoid the feedback between the propensity score model (i.e., treatment assignment model) and the intermediate/outcome models, a two-step Bayesian approach \citep{lunn2009combining, zigler2016central} could be employed. This approach allows for the prevention of feedback by decoupling the models and analyzing them in separate steps.

Further extension research covers a wide range of topics. If there are multiple intermediate variables, the intermediate model can be constructed using a multivariate response, and a multivariate principal stratum can be designed accordingly. In this case, the principal strata data entered the outcome model's modifier function as multivariate data. Considering the nature of air quality research, several variables (various weather variables, regional/geographical information, social information, etc.) can be considered as potential confounders where it is important to select the confounders that will actually be used. Using a recently proposed method \citep{kim2023bayesian}, it will be possible to connect a common selection probability prior to the prognostic function of each Bayesian causal forest model for the selection of the important confounders. Moreover, the outcome variable, ambient \PMTwo\ concentrations in the surrounding area, exhibits spatial correlation. Nevertheless, the covariate information employed in our application is anticipated to mitigate this residual spatial correlation. The Moran's I value (0.12, p-value 0.37), computed using an inverse distance weight matrix, also suggests an absence of remaining spatial dependence. However, in alternative applications, introducing a modification to include a spatial random effect may prove beneficial.

\section*{Acknowledgements}
The authors gratefully acknowledge that this work is supported by the National Research Foundation of Korea
(NRF) grant funded by the Korea government (NRF-2020R1F1A1A01048168, NRF-2022R1F1A1062904) and the US National Institutes of Health (R01ES034803 and R01ES035131). 

\section*{Data Availability Statement}
The data that support the findings of this paper are openly available in the GitHub repository at https://github.com/lit777/BPCF.

\bibliographystyle{asa_edit}
\bibliography{Bibliography}

\section*{Supporting Information}
Web Appendix A-E referenced in Sections 2, 3, 4 and 5 are available with this paper at the Biometrics website on Wiley Online Library. R code files for the simulation studies and data analysis are available with this paper at the Biometrics website on Wiley Online Library.

\end{document}